\documentclass[sn-nature]{sn-jnl}                   % Style for submissions to Nature Portfolio journals
\usepackage{graphicx}%
\usepackage{multirow}%
\usepackage{amsmath,amssymb,amsfonts}%
\usepackage{amsthm}%
\usepackage{mathrsfs}%
\usepackage[title]{appendix}%
\usepackage{xcolor}%
\usepackage{textcomp}%
\usepackage{manyfoot}%
\usepackage{booktabs}%
\usepackage{algorithm}%
\usepackage{algorithmicx}%
\usepackage{algpseudocode}%
\usepackage{listings}%
\theoremstyle{thmstyleone}%
\theoremstyle{thmstyletwo}%

\theoremstyle{thmstylethree}%

\begin{document}

% \title[Article Title]{Surrogate data-based machine learning for early warning signals of critical transitions}
% \title[Article Title]{Early warning signals of critical transitions in complex systems: Machine  learning via surrogate data}
% \title[Article Title]{Predicting critical transitions in complex systems using machine learning via surrogate data}
% \title[Article Title]{Predicting critical transitions with surrogate data-based machine learning}
\title[Article Title]{Learning from the past: predicting critical transitions with machine learning trained on surrogates of historical data}


\author[1]{\fnm{Zhiqin} \sur{Ma}}
%\author[1]{\fnm{Zhiqin} \sur{Ma}}\email{mzq\_2021@outlook.com}

\author*[1]{\fnm{Chunhua} \sur{Zeng}}\email{zchh2009@126.com}

%\author[1]{\fnm{Hui} \sur{Fang}}

%\author*[2,3]{\fnm{Ying-Cheng} \sur{Lai}}\email{Ying-Cheng.Lai@asu.edu}
%\author[2,3]{\fnm{Yi-Cheng} \sur{Zhang}}\email{yi-cheng.zhang@unifr.ch}
\author[2,3]{\fnm{Yi-Cheng} \sur{Zhang}}

\author*[4]{\fnm{Thomas M.} \sur{Bury}}\email{thomas.bury@mcgill.ca}

\affil*[1]{\orgdiv{Faculty of Science}, \orgname{Kunming University of Science and Technology}, \orgaddress{\city{Kunming}, \postcode{650500}, \country{China}}}
%\affil*[1]{Faculty of Science, Kunming University of Science and Technology, Kunming 650500, China}

%\affil[1]{\orgdiv{Department}, \orgname{Organization}, \orgaddress{\street{Street}, \city{City}, \postcode{10587}, \state{State}, \country{Country}}}
%
%\affil*[2]{\orgdiv{School of Electrical, Computer and Energy Engineering}, \orgname{Arizona State University}, \orgaddress{\city{Tempe}, \postcode{85287}, \state{Arizona}, \country{USA}}}
%\affil*[3]{\orgdiv{Department of Physics}, \orgname{Arizona State University}, \orgaddress{\city{Tempe}, \postcode{85287}, \state{Arizona}, \country{USA}}}

\affil[2]{\orgdiv{Department of Physics}, \orgname{University of Fribourg}, \orgaddress{\city{Fribourg}, \postcode{1700}, \country{Switzerland}}}
\affil[3]{\orgdiv{Swiss Center for Data and Network Sciences}, \orgname{University of Zurich}, \orgaddress{\city{Zurich}, \postcode{8050}, \country{Switzerland}}}

\affil*[4]{\orgdiv{Department of Physiology}, \orgname{McGill University}, \orgaddress{\street{3655 Promenade Sir William Osler}, \city{Montreal}, \country{Canada}}}


\abstract{Complex systems can undergo critical transitions, where slowly changing environmental conditions trigger a sudden shift to a new, potentially catastrophic state. Early warning signals for these events are crucial for decision-making in fields such as ecology, biology and climate science. Generic early warning signals motivated by dynamical systems theory have had mixed success on real noisy data. More recent studies found that deep learning classifiers trained on synthetic data could improve performance. However, neither of these methods take advantage of historical, system-specific data. Here, we introduce an approach that trains machine learning classifiers directly on surrogate data of past transitions, namely surrogate data-based machine learning (SDML). The approach provides early warning signals in empirical and experimental data from geology, climatology, sociology, and cardiology with higher sensitivity and specificity than two widely used generic early warning signals---variance and lag-1 autocorrelation. Since the approach is trained directly on surrogates of historical data, it is not bound by the restricting assumption of a local bifurcation like previous methods. This system-specific approach can contribute to improved early warning signals to help humans better prepare for or avoid undesirable critical transitions.}




\maketitle

\section{Introduction}\label{sec1}

% Intro to critical transitions
Many complex systems have critical thresholds called tipping points, at which an abrupt shift between qualitatively different states occurs~\cite{scheffer2020critical,van2016you,bi2024folding,panahi2023rate,murphy2024information}. In biology, critical transitions are associated with asthma attacks~\cite{sandberg2000role,venegas2005self}, epileptic seizures~\cite{jirsa2014nature,mcsharry2003prediction,kramer2012human}, microbiome dysregulation~\cite{lahti2014tipping,li2023role}, depression~\cite{beck2009depression,van2014critical} and cardiac arrhythmia~\cite{Bury2023Predicting,glass2022clocks}; in economics, financial markets can form a ``bubble'' and fall into a recession~\cite{may2008ecology,walker2011genealogies}; in climatology, abrupt shifts in ocean circulation or climate can occur~\cite{armstrong2022exceeding,ditlevsen2023warning,boers2021observation,brovkin2021past,lenton2008tipping}; and in ecology \cite{hastings2018transient,assessment2005ecosystems,scheffer2001catastrophic}, ecosystems can collapse due to their interplay with human behavior, resulting in catastrophic shifts in forests or grasslands~\cite{arani2021exit,smith2022empirical,henderson2016alternative}, eutrophication of lakes~\cite{ma2022relaxation,wang2012flickering,carpenter2005eutrophication}, collapse of populations~\cite{carpenter2011early,rigal2023farmland} and bleaching of coral reefs~\cite{hughes2017global,obura2022vulnerability}. These events are characterized by a sudden shift to a new, potentially catastrophic state and are referred to as critical transitions.

% Intro to generic early warning signals
Countless decisions made by individuals, industries, and policy-makers depend upon the accuracy of early warning signals for critical transitions. Examples range from deciding the rate at which to harvest a population of fish \cite{hutchings2000collapse} to deciding whether to flee a potentially dangerous earthquake \cite{sornette2002predictability}. In 2009, Scheffer and colleagues proposed monitoring the generic indicators variance and lag-1 autocorrelation to determine whether a critical transition is approaching \cite{scheffer2009early}. This was motivated by a universal phenomenon called critical slowing down, which occurs in the vicinity of local bifurcations in dynamical systems \cite{strogatz2018nonlinear, wissel1984universal,kuznetsov1998elements}. Critical slowing down is characterized by an increased return time to equilibrium following a perturbation, which, in noisy systems, is manifested as an increase in variance and lag-1 autocorrelation. Rising variance and lag-1 autocorrelation was subsequently found prior to critical transitions in complex systems in cardiology~\cite{quail2015predicting,Bury2023Predicting}, ecology~\cite{carpenter2011early,wang2012flickering},
the climate~\cite{dakos2008slowing,boers2018early,boers2021observation},
engineering~\cite{pavithran2021effect} and geology~\cite{hennekam2020early}, to name a few areas. Many limitations to these generic early warning signals also emerged. They can fail to signal a transition if the time series is too short, too noisy~\cite{perretti2012regime}, or too non-stationary~\cite{clements2016rate,ashwin2012tipping}, or if the transition corresponds not to a local bifurcation, but a global bifurcation~\cite{hastings2010regime}, or no bifurcation at all~\cite{ashwin2012tipping}. The development of more effective early warning signals remains a highly active area of research \cite{dakos2023tipping, ma2022relaxation}.

% ~\cite{Bury2023Predicting,ma2024relaxation}.

% Intro to deep learning methods for EWS
In recent years, deep learning classifiers have been employed for early warning signals~\cite{Bury2023Predicting,bury2021deep,ismail2019deep,deb2022machine,dylewsky2023universal}. These classifiers require a lot of data, which are typically not available in the context of critical transitions. Studies have therefore opted to use synthetic data generated from simulations of mathematical models going through bifurcations---a dataset that can be made as large as needs be~\cite{bury2021deep}. Given the universal properties of bifurcations~\cite{kuznetsov1998elements,strogatz2018nonlinear}, these classifiers have been shown to be effective at predicting bifurcations in real systems, outperforming variance and lag-1 autocorrelation in many cases~\cite{bury2021deep,Bury2023Predicting,deb2022machine}. However, machine learning classifiers are only as good as their training data---this approach is reliant upon the real-world system exhibiting similar properties to one of the bifurcation trajectories contained within the training set. Indeed, it makes the assumption that a bifurcation exists in the first place---something that is not inherent to real-world systems, but rather a property of a mathematical model for the system of interest~\cite{strogatz2018nonlinear}.

% why surrogate methods
Ideally, a machine learning classifier would be trained directly on data of the real-world system. To get around the problem of limited data availability, we propose using surrogate data methods to generate trajectories derived from data of past critical transitions. This way, we can generate a lot of training data (enough to train a machine learning classifier) while also preserving system-specific properties of the data from past critical transitions.

\begin{figure}[ht]%
\centering
\includegraphics[width=1.0\textwidth]{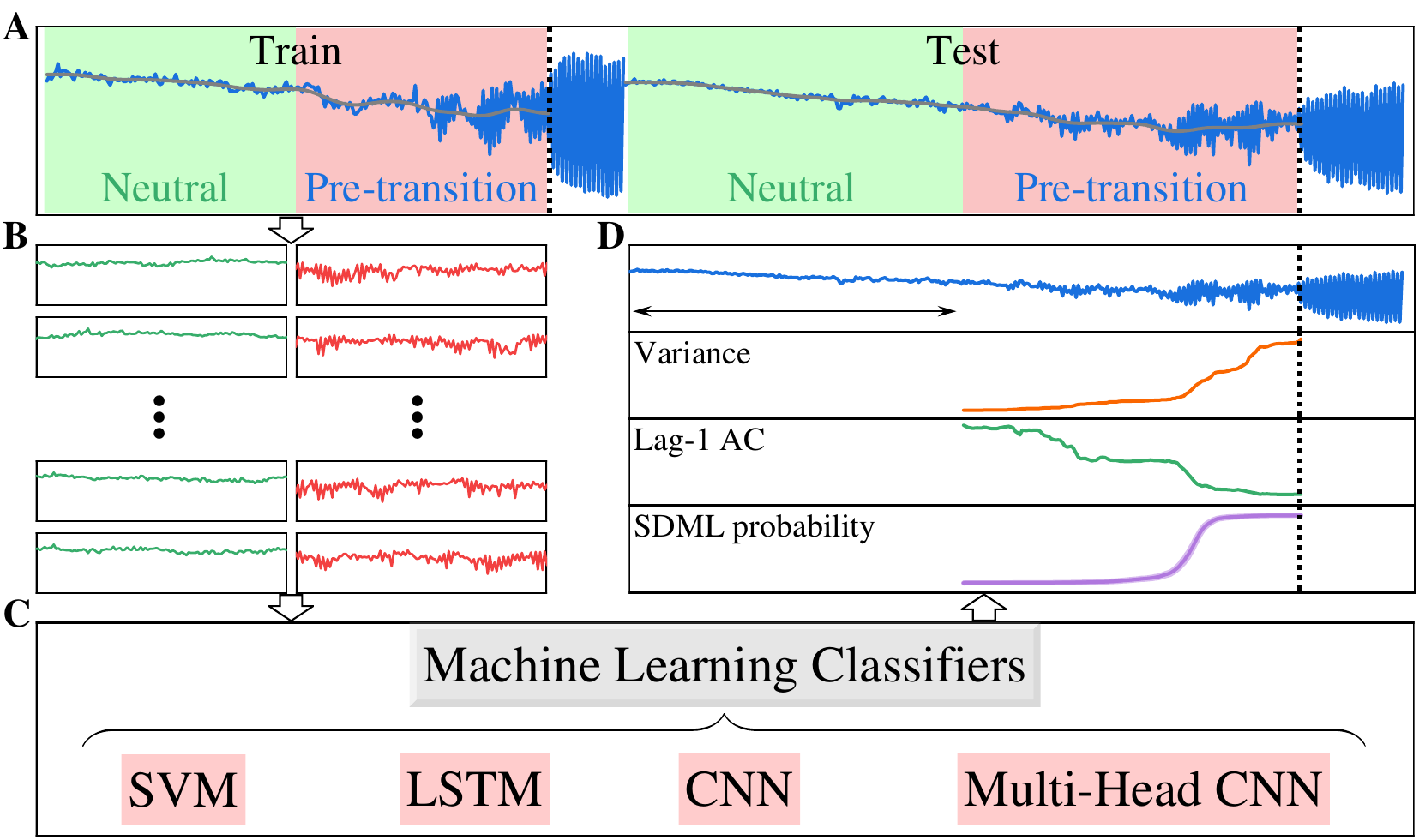}
\caption{\textbf{Illustration of the SDML prediction framework.} (\textbf{A}) Two trajectories (blue) and smoothing (gray) of chick heart aggregates approaching a critical transition. The vertical dashed line marks the onset of the transition. A green background denotes sections of the time series taken as far from the transition (``Neutral") and red denotes sections close to the transition (``Pre-transition"). The trajectories are divided into training (left) and test (right) trajectories. (\textbf{B}) Thousands of surrogate time series are generated from the neutral (left) and pre-transition (right) training trajectories. (\textbf{C}) Machine learning classifiers used for the binary classification problem of distinguishing neutral from pre-transition time series. We use support vector machines (SVM), long short-term memory (LSTM) networks, convolutional neural networks (CNN), and Multi-Head CNN. (\textbf{D}) The changing trends of indicators prior to the transition in the test trajectory, including variance, lag-1 autocorrelation (AC), and probabilities assigned by the SDML classifier. The arrow illustrates the rolling window (50\% of the time series) used for computing early warning signals.}
\label{figure_1}
\end{figure}

% Background on surrogate methods
Surrogate data methods have been primarily developed to enable robust statistical evaluations about a system from limited data~\cite{lancaster2018surrogate,theiler1992testing,schreiber1996improved}. The method involves creating multiple versions of a time series with the same (or very similar) statistical properties, including the histogram distribution, power spectrum and autocorrelation function~\cite{birkholz2015predictability,lancaster2018surrogate}. This allows one to test against a null hypothesis using a larger number of samples. Here, we use surrogate data methods to boost the number of samples with which to train machine learning classifiers. Specifically, surrogate data is generated from data far from (neutral) and near to (pre-transition) the critical transitions, serving as two distinct classes (Fig.~\ref{figure_1}). We then train machine learning classifiers to distinguish between these two classes. When a classifier assigns a high probability to the pre-transition class, this is interpreted as an early warning signal for a critical transition.

% Summary of what we do
This study proposes the surrogate data-based machine learning (SDML) approach for early warning signals of critical transitions. We test the SDML approach on empirical data from geology, climate, sociology and cardiology, and compare its performance to the widely used generic early warning signals, variance and lag-1 autocorrelation. We also assess the robustness of SDML approach to different surrogate data methods.

% reduce the demand for high-cost data collection.

% capturing features (or patterns) in data that are not easily represented in bifurcations.

% Each SDML classifier was trained on many thousands of surrogate data samples generated from system-specific historical data.

% makes no assumptions about a bifurcation and simply looks for patterns specific to a particular system that may (or may not) occur prior to the critical transition.

% Namely, the approach does not need to know ahead of time what type of bifurcations that the real systems belong to.

% the ability of surrogate datasets to be used as training data in machine learning classifiers has not been investigated yet.

 %%% These should be in results/discussion

 % The comparison results indicate that the SDML has higher sensitivity and specificity than commonly used early warning signals, such as variance and lag-1 autocorrelation. In addition, the out-of-sample predictive performance is evaluated experimentally using five different surrogate data methods.

 % The experimental results demonstrate that the SDML achieve robust prediction performance using different surrogate data methods. Therefore, inspired by successful applications in surrogate data, we believe that the SDML approach is a promising solution to capture the underlying patterns of critical transitions, allowing us to achieve the early warning signals tailored to a specific system.

\section{Results}\label{sec2}

% \subsection{Frameworks of EWS on chick heart data}\label{subsec2_1}
% \subsection{Framework of EWS for surrogate data-based machine learning}\label{subsec2_1}

\subsection{Rapid transition events}\label{subsec2_1}

\begin{figure}[ht]%
\centering
\includegraphics[width=0.8\textwidth]{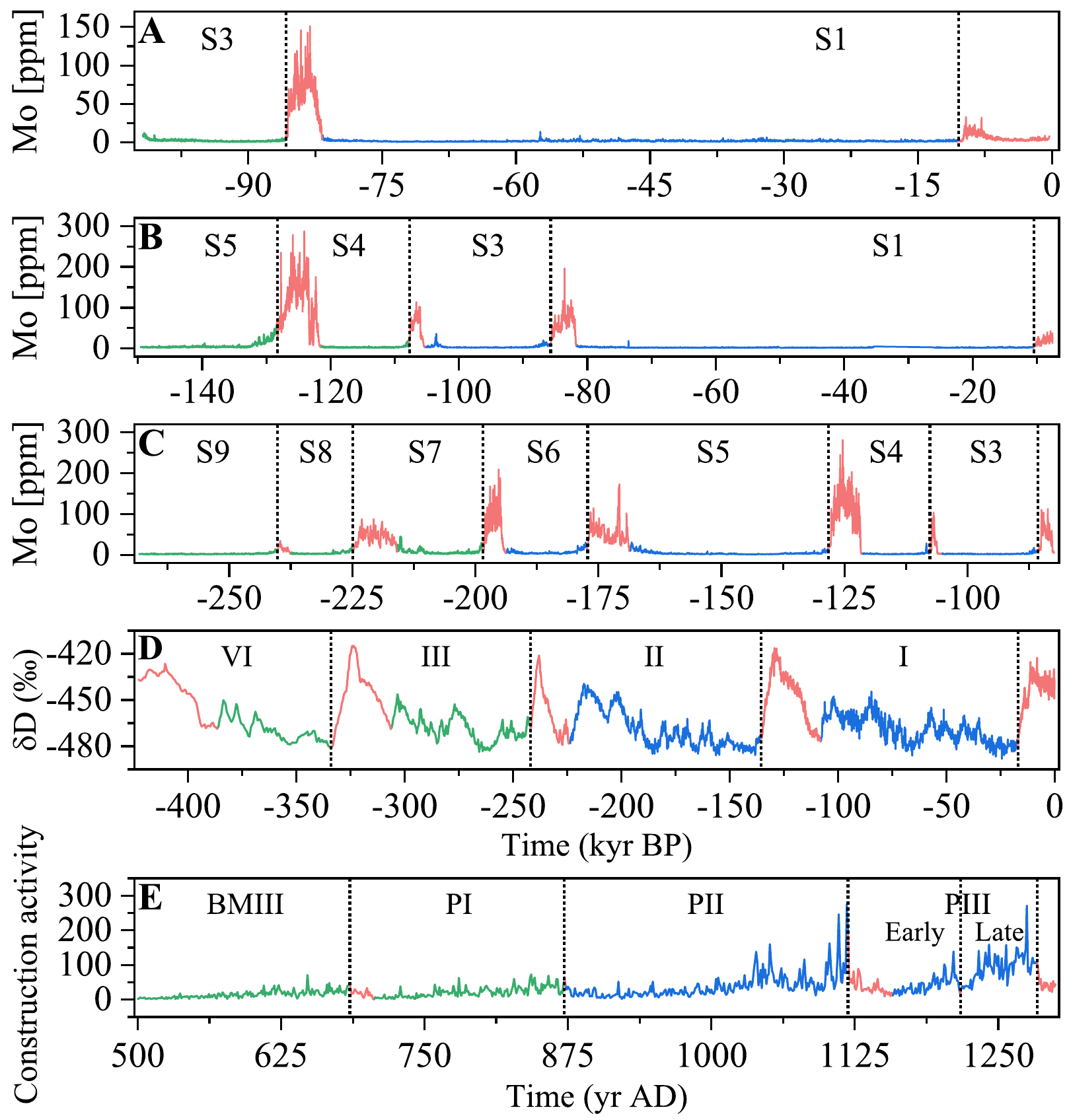}
\caption{\textbf{Time series of rapid transition events for consecutive recordings from three different empirical systems.} (\textbf{A} to \textbf{C}) Sedimentary archives from the Mediterranean Sea for core MS21 at depth 1,022 m (A), core MS66 at depth 1,630 m (B), and core 64PE406E1 at depth 1,760 m (C). (\textbf{D}) Paleoclimate transitions from Deuterium content expressed as $\delta$D (in $\text{\textperthousand}$ with respect to the standard mean ocean water). Time is given as years before present (BP). \textbf{(E}) Construction activity in the pre-Hispanic Pueblo societies, given as the number of trees felled per year. Vertical dashed lines indicate onsets of rapid transition events. Green lines are historical trajectories used to generate  the surrogate data. Red lines are post-transition trajectories, not used. Blue lines are trajectories used for testing the early warning signals.}
\label{figure_2}
\end{figure}

% Description of rapid transition events and how data used
Figure~\ref{figure_2} shows rapid transition events in consecutive recordings of three empirical systems: anoxic transitions in the Mediterranean Sea~\cite{hennekam2020early}, paleoclimate transitions~\cite{petit1999climate}, and transitions in human activity in the pre-Hispanic Pueblo societies~\cite{scheffer2021loss}. Surrogate data is generated from historic pre-transition data, shown in green in Fig.~\ref{figure_2}. These data are used to predict future periods, shown in blue. The SDML approach to predicting critical transitions is motivated by two hypotheses: (i) the trajectories leading up to historic and future critical transitions have similar dynamical features; (ii) trajectories far from a critical transition (denoted ``neutral") have different dynamical features to trajectories close to a critical transition (denoted ``pre-transition''). If these hypotheses hold, it should be possible to train a machine learning classifier to predict future critical transitions using SDML.

% Anoxic data
Due to the increasing temperatures and input of excess nutrients to coastal regions, the oceans and seas are losing oxygen. In the geological past, large‐scale anoxic events had occasionally triggered mass extinction events, and the current loss of oxygen in the marine field has increased concerns about the recurrence of these types of events. To understand the marine anoxic events, Hennekam et al.~\cite{hennekam2020early} reconstructed past oxygen conditions in the Mediterranean Sea using redox‐sensitive trace elements. The high‐resolution anoxia records, based on Mo, indicated a rapid oxic‐to‐anoxic transition at the start of each sapropel, as shown in Figs. \ref{figure_2}a--\ref{figure_2}c. Also, rapid oxic‐to‐anoxic transitions occurred regularly in the eastern Mediterranean Sea on (multi)centennial time scales. In addition, deoxygenation events in the eastern Mediterranean Sea showed tipping point behavior during which the system abruptly shifted to an anoxic state.

% Ice core data
The completion of drilling at Vostok station in East Antarctica has allowed the extension of the ice record of atmospheric composition and climate to the past four glacial-interglacial cycles~\cite{petit1999climate}. Glacial–interglacial climate changes have been documented by complementary climate records derived from deep-sea sediments, continental deposits of flora, fauna and loess, and ice cores. The Vostok ice-core record extended through four climate cycles, with the ice slightly older than 400~kyr at a depth of 3,310~m, covering a period comparable to that covered by numerous oceanic and continental records. Figure~\ref{figure_2}d shows the deuterium content of the ice (i.e., a proxy of local temperature change) as a function of the GT4 timescale of the ice. A prominent feature of the glacial-interglacial cycles was the abrupt termination of most glacial periods.

% Socio data
The collapse of civilizations has remained one of the most enigmatic phenomena in human history. To explore the collapse of ancient societies, Scheffer et al.~\cite{scheffer2021loss} leveraged an extraordinarily long and high-resolution time series of Puebloan tree-cutting as a proxy for construction, shown in Fig.~\ref{figure_2}e. The time series spanned eight centuries and five transitions. The annual-resolution time series of the construction activity demonstrated that repeated dramatic transformations of Pueblo cultures in the pre-Hispanic US Southwest were preceded by signals of critical slowing down.

\subsection{Performance of early warning signals on experimental and empirical data}\label{subsec2_2}

\begin{figure}[ht]%
\centering
\includegraphics[width=1.0\textwidth]{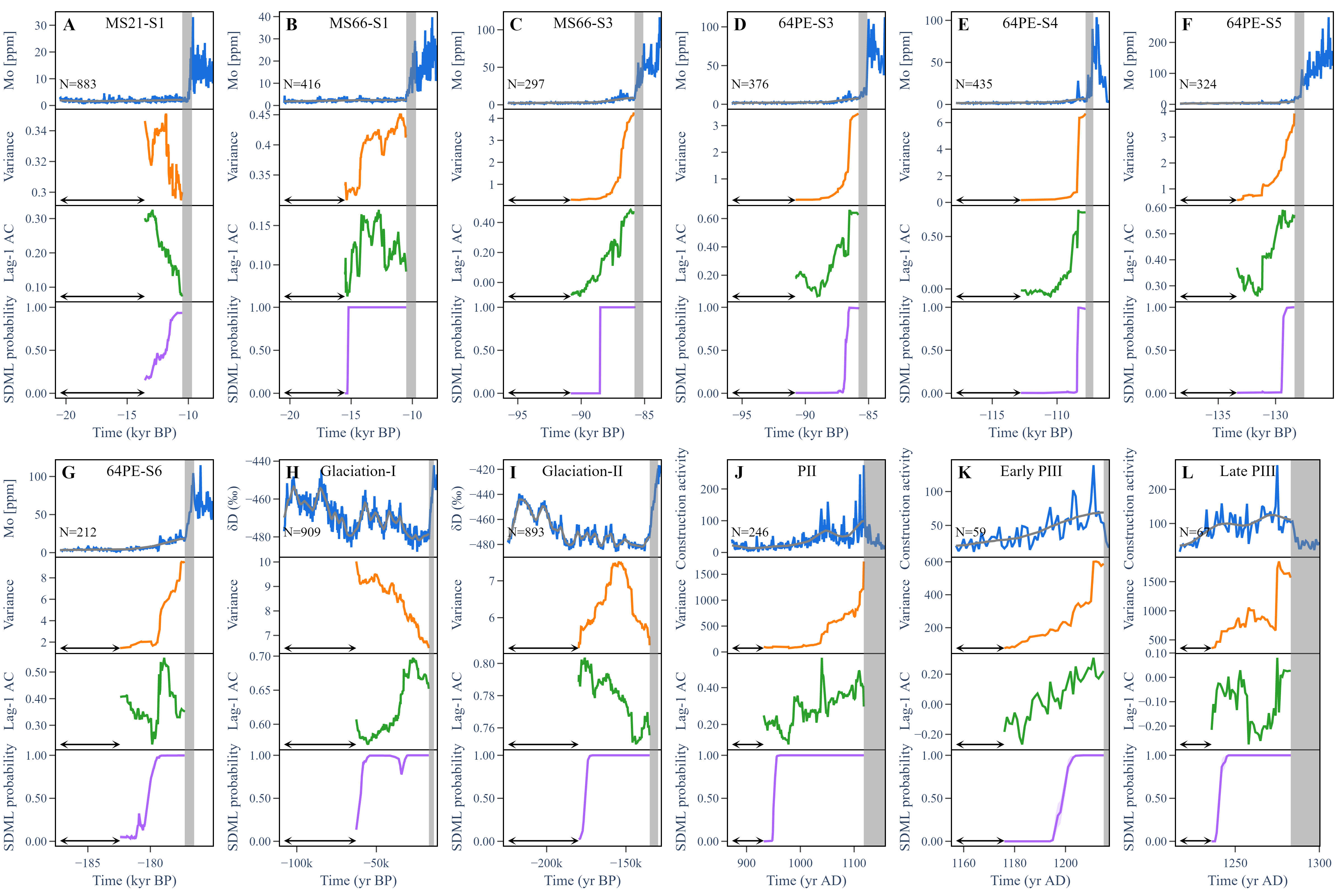}
\caption{\textbf{Trends in indicators prior to rapid transition events in empirical data using the amplitude adjusted Fourier transform (AAFT) surrogate method.} Trajectories correspond to blue traces in Fig. 2. (\textbf{A}) Anoxic transition (Sapropel S1) obtained from data on the MS21 core. (\textbf{B--C}) Anoxic transitions (Sapropels S1 and S3) obtained from data on the MS66 core. (\textbf{D--G}) Anoxic transitions (Sapropels S3 to S6) obtained from data on the 64PE406E1 core. (\textbf{H}) End of glaciation I (i.e., the end of the last glaciation). (\textbf{I}) End of glaciation II. (\textbf{J}) Archaeological Period PII. (\textbf{K}) Archaeological Period Early PIII. (\textbf{L}) Archaeological Period Late PIII. (Top) Trajectory (blue) and Gaussian smoothing (grey); (Second down) Variance; (Third down) Lag-1  autocorrelation (AC); (Bottom) probability of an approaching critical transition assigned by the SDML classifier. The lines and shaded areas show the mean and 95\% confidence interval, respectively. The arrows indicate the width of the rolling window used to compute early warning signals. The grey bands show transition phases.} \label{figure_3_AAFT}
\end{figure}

We trained and tested the SDML classifier on empirical and experimental data from ecology, climatology, sociology, and cardiology, compared its performance with variance and lag-1 autocorrelation. The indicators were monitored progressively as more of the time series was revealed. Variance and lag-1 autocorrelation were considered to provide an early warning signal if they displayed a strong trend, which was quantified using the Kendall tau statistic. For variance, an increasing trend was taken as a warning signal. For lag-1 autocorrelation, the direction of the trend depends on the frequency of oscillations ($\theta$) at the transition. For $\theta \in [0, \pi/2)$, the lag-1 autocorrelation is expected to increase, whereas for $\theta \in (\pi/2, \pi]$, it is expected to decrease~\cite{bury2020detecting}. In the chick heart aggregates, an increase in variance and a decrease in lag-1 autocorrelation is observed prior to the transition \cite{quail2015predicting}, which could serve as an early warning signal (Fig.~\ref{figure_1}d). For the test trajectories in the three empirical datasets (Fig.~\ref{figure_2}), blue), variance and lag-1 autocorrelation showed mixed success at providing an early warning signal. Variance increased prior to the transition in nine out of the twelve cases (Fig.~\ref{figure_3_AAFT}, orange) and lag-1 autocorrelation increased in seven (Fig.~\ref{figure_3_AAFT}, green).

The SDML classifier assigns a probability to the two possible classes: neutral (far from the transition) and pre-transition (close to the transition). An increase in the probability for the pre-transition class is taken as an early warning signal (Fig.~\ref{figure_3_AAFT}, purple). For each case, the probability of a pre-transition state increases over time, thereby providing an early warning signal. Close to the transitions, the classifier becomes very confident of a pre-transition state, with a probability at or close to one. These results provide supporting evidence for the hypotheses outlined in this study. Moreover, the SDML approach was effective for five different surrogate data methods (Figs. S1--S5), demonstrating its robustness to this choice.

\begin{figure}[ht]%
\centering
\includegraphics[width=1.0\textwidth]{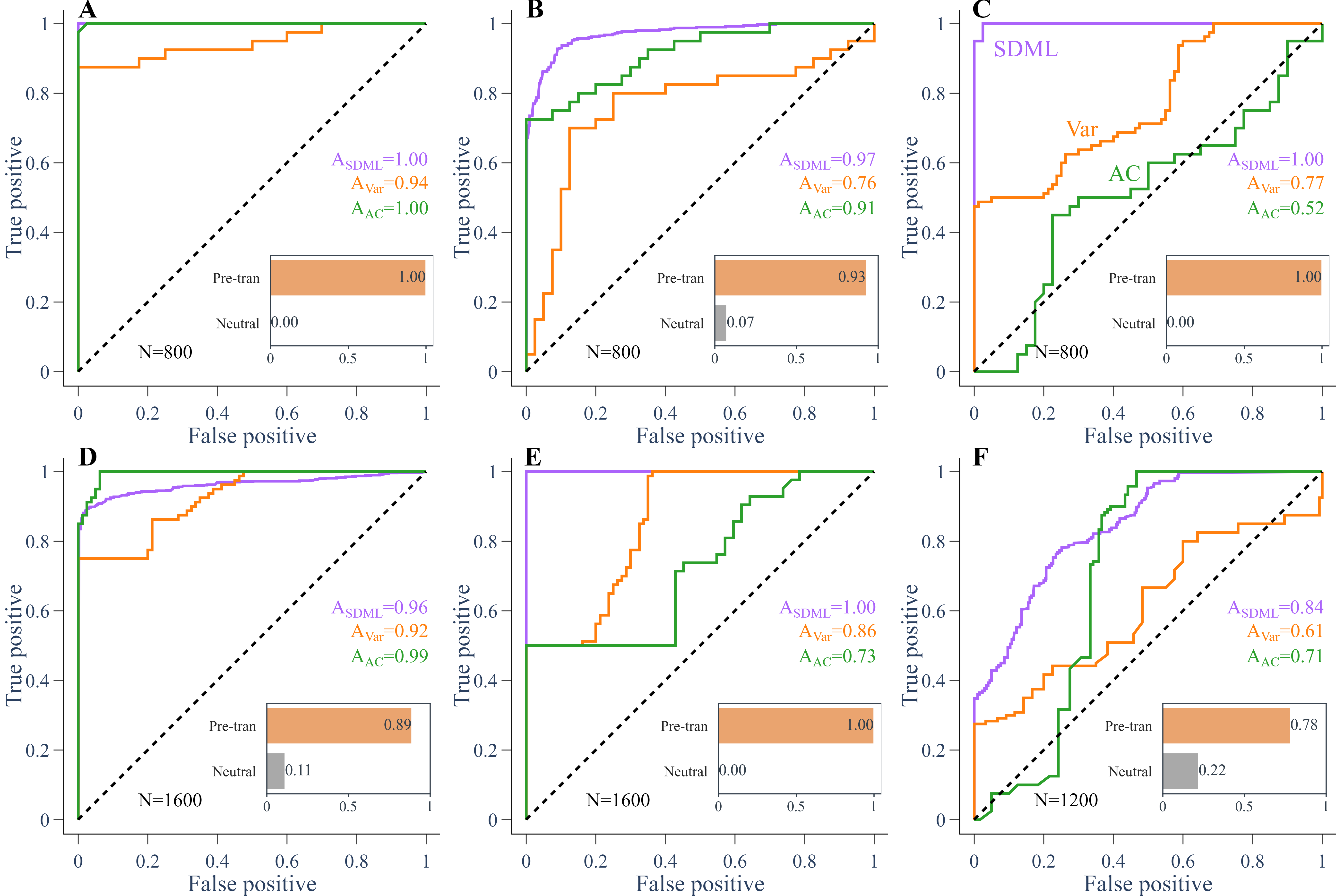}
\caption{\textbf{ROC curves showing performance of indicators in the experimental and empirical data.} The ROC curves show the SDML classifier (SDML, purple), variance (Var, orange), and lag-1 autocorrelation (AC, green) for the (\textbf{A}) chick heart aggregates going through a period-doubling bifurcation; (\textbf{B}) sediment data from the MS21 core, (\textbf{C}) MS66 core, and (\textbf{D}) 64PE406E1 core showing rapid transitions to an anoxic state in the Mediterranean Sea; (\textbf{E}) ice core records showing rapid paleoclimate transitions; and (\textbf{F}) transitions in construction activity in pre-Hispanic Pueblo societies. Predictions were obtained from 10 classifiers for each dataset, with 40 equally spaced predictions made between 60\% and 100\% (a, b, e, and f) or between 80\% and 100\% (c and d) of the way through the pre-transition data. The area under the curve (AUC), denoted by A, is a performance measure. The insets show the proportion of predictions made by the classifier for true pre-transition trajectories. ``Pre-tran" means close to a critical transition, and ``Neutral" means far from a critical transition.}
\label{figure_4_AAFT}
\end{figure}

These time series, however, do not address how the indicators might perform when faced with a neutral time series where no transition occurs, and whether they might mistakenly generate a false positive prediction of an oncoming state transition~\cite{bury2021deep,boettiger2012early}. Therefore, we compare the performance of these approaches with respect to both true and false positives using the receiver operator characteristics (ROC) curve. The ROC curve shows the ratio of true positives to false positives as a discrimination threshold is varied. For variance and lag-1 autocorrelation, the discrimination threshold comes from the Kendall tau value. For the SDML classifiers, it comes from the classifiers assigned probability of a pre-transition state. The area under the ROC curve is a measure of performance based on both sensitivity (how many true positives were detected) and specificity (how many false positives were avoided). An AUC of 1.0 indicates a perfect classifier, and an AUC value of 0.5 indicates a classifier that is no better than random.

To construct the ROC curves, we make predictions on both ``pre-transition" time series (close to the critical transition) and ``neutral" time series (far the critical transition). The ROC curves for variance, lag-1 AC and the SDML classifier for six evaluations across four study systems are shown in Fig.~\ref{figure_4_AAFT}. For the chick heart data (Fig.~\ref{figure_4_AAFT}a), the SDML classifier and lag-1 AC equally outperformed variance. For the other empirical datasets, the SDML classifier outperformed both variance and lag-1 autocorrelation in all cases except for the sediment data of the 64PE406E1 core (Fig.~\ref{figure_4_AAFT}d), where lag-1 autocorrelation was the best indicator.

\section{Discussion}\label{sec12}

% summary and perspective
Many complex systems are prone to critical transitions and data from past transitions are often limited. Nevertheless, data from past transitions contain important system-specific information that should not be neglected. The majority of research on early warning signals has focused on generic indicators based on properties of bifurcations, and has not incorporated system-specific details, nor information from past transitions. In this paper, we have shown that the limited data available from past transitions can be extended using surrogate data methods, to the point where a large enough dataset can be generated to train a machine learning classifier. These trained machine learning classifiers are capable of providing early warning signals for unseen, future critical transitions, despite having only been trained on surrogate data. Moreover, we found that the classifiers outperformed variance and lag-1 autocorrelation in most experimental and empirical test cases among natural systems in cardiology, geology, climatology and sociology.

% advantage over generic indicators
The results suggest that the SDML method for early warning signals can be advantageous over generic indicators. There are many scenarios for which generic early warning signals can fail~\cite{dakos2015resilience}. Examples include global bifurcations~\cite{hastings2010regime}, out-of equillibrium (transient) dynamics~\cite{hastings2018transient}, and rate-induced tipping~\cite{ashwin2012tipping}. These problems stem from generic early warning signals relying on a manifestation of critical slowing down. In contrast, the SDML method makes no assumption of a bifurcation---it simply observes past transitions and attempts to extract features that are associated with the approach of a transition. For cases such as the onset of alternans in chick heart aggregates, where a period-doubling bifurcation is the likely cause, it seems plausible that the classifier learns about critical slowing down through observing changes in variance and lag-1 autocorrelation. However, for cases where a bifurcation mechanism is not clear, such as for the paleoclimate transitions, learning system-specific features could be key to an effective early warning signal.

%Common cases include time series that are too short, too noisy, or too non-stationary, or transitions arising not from local bifurcations, but global bifurcations, or no bifurcation at all.

%In cases such as the chick heart data, where there is a clear bifurcation, the classifiers presumably learns about critical slowing down through observing different variance and  lag-1 autocorrelation in the pre-transition sections, though importantly it can also learn about features that are system specific--features that may not be expressed as an increase in variance and lag-1 autocorrelation.

% Tuning hyperparameters / new architectures to achieve better performance
Machine learning is a rapidly evolving field, with new algorithms and architectures appearing every year. It is possible that more recent machine learning architectures (e.g. transformers~\cite{ashish2017attention}) could achieve a higher performance than the results presented here, and we have not attempted a comprehensive assessment of the performance of all appropriate types of architecture. We have also not performed hyperparameter tuning on the individual classifiers. Rather, this study is focused on the presentation of a new framework and demonstration of its baseline capabilities, that can be fine-tuned in future work.

 % It should be noted that this study primarily focused on the ability to use historical data-generated surrogate datasets as training data for machine learning classifiers. Therefore, only the baseline algorithms of the machine learning classifiers were considered without tuning the classifiers' hyperparameters to optimize the performance. The reason why different machine learning baseline algorithms were used for the SDML classifier was that no single algorithm could necessarily achieve the best performance in different fields of real-system prediction. Therefore, this study selected the best machine learning baseline algorithm for each rapid transition event. Generally, the machine learning classifiers used in this study could be regarded as a family of models, with the current version being only a baseline algorithm, having the potential to be further extended to improve prediction accuracy in the future. In addition, a machine learning classifier for time series classification, such as transformers~\cite{ashish2017attention}, could be designed to achieve higher performance on the test data used in this study.

An important consideration in this approach is how to build the training library for the machine learning classifiers. We opted for a simple approach of creating two classes of time series---one containing data far from the critical transition, and the other containing data close to it. More elaborate schemes could be considered in future work, such as having several classes for data at different distances to the transition, or turning the problem into a regression task, where time series are labelled with their distance to the transition.

The data-driven SDML approach relies on good quality data of past transitions. This is in order to generate high fidelity surrogate data for the machine learning classifiers. Therefore, rich, high-quality data sources with rapid transition events (e.g.,~sedimentary archives from the Mediterranean Sea~\cite{hennekam2020early}, paleoclimate transitions~\cite{petit1999climate,dakos2008slowing} and construction activity in pre-Hispanic Pueblo societies~\cite{scheffer2021loss}) and experimental data~(e.g., embryonic chick heart cell aggregates~\cite{Bury2023Predicting} and thermoacoustic instability experiments~\cite{pavithran2021effect}), are invaluable. The SDML approach should not be regarded as a replacement for generic early warning signals, which have been used and enhanced for decades and rigorously tested in many real-world contexts. Rather, our work should be interpreted as evidence that the proposed approach is able to meet the challenges of real-world forecasting problems and have the potential to complement and improve the current best methods.

The SDML approach opens a new direction for research on early warning signals, with applications in ecology, climate, finance, energy, and human and biological activities, as well as other complex dynamical systems. The authors believe that machine learning classifiers trained on rich surrogate data of past transitions could be crucial in advancing our ability to prepare for or avert critical transitions.

\section{Methods}\label{sec_Methods}
%Topical subheadings are allowed. Authors must ensure that their Methods section includes adequate experimental and characterization data necessary for others in the field to reproduce their work. Authors are encouraged to include RIIDs where appropriate.

\subsection{Training data generation}\label{sec_Methods_1}

The training data consisted of thousands of surrogate data samples, which were generated from data close to and far from critical transitions in the empirical systems. Time series between critical transitions were split at their mid-point with the earlier half taken as ``neutral" and the later half taken as ``pre-transition". Surrogate data were generated for the ``neutral" and ``pre-transition" sections using a variety of methods, which preserve a the histogram distribution and/or the power spectrum of the original data. The surrogate data were labelled according to whether they were far from (neutral) or close to (pre-transition) a critical transition, making up a binary categorization dataset. The methods used to generate surrogate data are listed below.

\textbf{Random permutation (RP) surrogates:} RP surrogates are generated by randomly shuffling the original time series. They possess the same mean, variance, and histogram distribution as the original signal, but any temporal structure is removed~\cite{faes2004surrogate}. They therefore possess a different power spectrum and autocorrelation function. RP surrogates were developed to test for temporal structure in a signal.

\textbf{Fourier transform (FT) surrogates:} In contrast to RP surrogates, FT surrogates preserve the power spectrum of the original signal, but not necessarily the histogram distribution. They are generated by taking the Discrete Fourier transform of the original signal, randomizing the phases, and transforming back into the time domain. They remove nonlinear dependencies while retaining all linear information, and so are often used to test for nonlinear structure in a signal.

\textbf{Amplitude adjusted Fourier transform (AAFT) surrogates:} AAFT surrogates aim to preserve both the power spectrum and the histogram distribution. They are generated similarly to FT surrogates, except a histogram transformation is applied in order match the histogram of the surrogate with the histogram of the original signal~\cite{paluvs1995testing,theiler1992testing}. However, this process can introduce spurious harmonics in the power spectrum. Therefore AAFT surrogates perfectly preserve the histogram distribution but the power spectrum is slightly distorted.

\textbf{Iterative amplitude adjusted Fourier transform (IAAFT1 and IAAFT2) surrogates:} IAAFT surrogates were introduced by Schreiber \& Schmitz~\cite{schreiber1996improved} to overcome the distortion of the power spectrum in AAFT surrogates. They are generated by iterative replacement of Fourier amplitudes to achieve a closer match between both the histogram distribution and the power spectrum. However, the IAAFT algorithm cannot preserve both the histogram distribution and the power spectrum exactly. There are two versions: IAAFT1, which exactly preserves the histogram distribution, and IAAFT2 which exactly preserves the power spectrum.

\subsection{Machine learning algorithms}\label{sec_Methods_2}

Machine learning algorithms used in this study include support vector machine (SVM), long short-term memory (LSTM), convolutional neural network (CNN), and Multi-Head CNN. We experimented with different algorithms since no single algorithm should necessarily achieve the best prediction performance across different fields and real systems. In this study, we select the best-performing machine learning algorithm for each type of rapid transition event.
% Additional machine learning algorithms could also be used with the SDML framework.

The best performing classifier for the MS66 sediment core and the ice core was the SVM, with F1 scores of 1.0 and 1.0, respectively. For the 64PE406E1 sediment core it was the LSTM, with an F1 score of 0.99. For the MS21 sediment core, it was the CNN with an F1 score of 0.99. Finally, for the chick heart data and the construction activity, it was the Multi-Head CNN, with F1 scores of 1.0 and 0.99, respectively. Implementation details of each algorithm are provided below. In each case, we use default hyperparameter values without hyperparameter tuning.

1) SVM: This is a supervised machine learning algorithm suitable for classification tasks in high-dimensional spaces. SVM works by finding the optimal hyperplane that separates the different classes in feature space. For time series classification, each data point is used as a feature.

% The hyperplane dimension depends on the feature number; namely, if the number of input features is two, then the hyperplane represents a line. In this study, the number of input features is two, so the hyperplane is a line.

2) LSTM: This is a type of recurrent neural network (RNN) that aims at handling the vanishing gradient problem present in traditional RNNs. The LSTM is effective at capturing temporal dependencies in sequential data, making it well suited to time series classification. We combine the LSTM with fully connected input and output layers. Specifically, we use a dense layer from the input time series to an LSTM layer of 128 neurons. To mitigate overfitting, we add a dropout layer with a dropout rate of 0.5, which was followed by another LSTM layer. Then we add a dense layer with 128 neurons and the ReLU activation function. Finally, we add a dense layer with the sigmoid activation function to output a probability distribution over the two classes.

3) CNN: This is a type of feed-forward neural network that learns features through convolution and can be used for time series classification. We use a CNN with two 1D convolutional layers with 64 filters, a kernel size of three and the ReLU activation function. They are followed by a dropout layer with a rate of 0.5 and a MaxPooling layer to reduce the dimension. Then, we use a dense layer with the ReLU activation function and 128 neurons. Finally we use a dense layer to the two output classes with a sigmoid activation function to output a probability distribution.

4) Multi-Head CNN: This algorithm was designed to capture features in the input data occurring on different scales. It does this through having distinct CNN heads, each having a different kernel size. We use three CNN heads. The first is a 1D convolutional layer with 32 filters, a kernel size of three, batch normalization to standardize inputs, and the ReLU activation function. This is followed by a dropout layer for regularization, a MaxPooling layer to reduce the dimensionality, and flattening. The second head is similar, but uses a kernel size of five, aiming to capture patterns over a broader window of input data. It is followed the same sequence of batch normalization, ReLU activation, dropout, max pooling, and flattening blocks. The third head uses a larger kernel size of 11 and 64 filters, designed to capture even broader patterns. The flattened outputs from all three heads are concatenated into a single vector, and then passed through a dense layer with 128 neurons (ReLU activation) and, finally, through the dense output layer with two neurons and sigmoid activation to output a probability distribution.

The LSTM, CNN, and Multi-Head CNN algorithms employed binary cross entropy as a loss function, which is convenient for binary classification problems, and the Adam optimizer for efficient network weight update. They are then trained using a batch size of 64 for 100 epochs. We use a training/validation split of 0.8/0.2. We used early stopping to prevent overfitting. This terminates the training when the validation accuracy does not improve for more than five epochs. Subsequently, we use the model checkpoint that achieved the highest validation accuracy. We report the F1 score of the model's performance on the validation set, which is a measure that takes into account both sensitivity and specificity. We note that the empirical test data used to generate the ROC curves is not involved in the training process.

\subsection{Experimental and empirical data used for testing}\label{sec_Methods_3}

\textbf{Embryonic chick heart cell aggregates:} We used publicly available data from experiments of chick heart cell aggregates that transition to a caridac arrhythmia \cite{Bury2023Predicting,quail2015predicting}. The aggregates were treated with 0.5~$\mu$mol--2.5~$\mu$mol of E4031, which is a drug that blocks the human Ether-$\grave{a}$-go-go-Related Gene (hERG) potassium
channel~\cite{clay1994review}. Following administration of the drug, in some aggregates, the time interval between two beats began to alternate; namely, there
was a period doubling bifurcation. Such transitions can also occur in the human heart in the form of T-wave alternans, which increase a patient’s risk for sudden cardiac death. The inter-beat intervals were computed as a time between consecutive beats and used in the analysis. The onset of the period-doubling bifurcation has been defined by Bury et al.~\cite{Bury2023Predicting} as the first time when the slope of a linear regression of the return map composed of a sliding window of inter-beat intervals is below -0.95 for the next 10 beats. There are 23 time series of aggregates going through a period-doubling bifurcation, and 23 time series from aggregates that did undergo any form of transition. Data were preprocessed as in Bury et al. \cite{Bury2023Predicting}, using a Guassian filter with a bandwidth of 20 beats. Early warning signals were computed using a rolling window of 0.5. In this study, for smoothing and computing generic early warning signals, we used the Python package ewstools \cite{bury2023ewstools}.

\textbf{Sedimentary archives from the Mediterranean Sea:} Hennekam et al.~\cite{sedimentary_archives, hennekam2020early} reconstructed past oxygen dynamics for three sediment cores (MS21, MS66, 64PE406E1) in the eastern Mediterranean Sea. Rapid transitions between oxic and anoxic states occurred regularly in this region in the geological past. Core MS21 spans 2 anoxic events, core MS66 spans 4 anoxic events, and core 64PE406E1 spans 7 anoxic events. The SDML approach was tested on each core separately. The sampling rate provided $\sim$10 to 50 years resolution depending on the core, with almost regular spacing between data points. This study performed the same data preprocessing as Hennekam et al.~\cite{hennekam2020early}, where residuals were obtained from smoothing the data with a Gaussian kernel with a bandwidth of 900 years, and early warning signals were computed using a rolling window of 0.5.

\textbf{Ice core records of paleoclimate transitions:} We used data from the Antarctica Vostok ice core which captures paleoclimate transitions from 420,000 years ago up to present day ~\cite{petit1999climate,Petit_Vostok,dakos2008slowing}. Out of the eight paleoclimate transitions analyzed by Dakos et al. \cite{dakos2008slowing}, we only consider the four transitions contained within the Vostok ice core since this study requires consecutive transitions contained within a single record. The data were preprocessed as in Dakos et al.~\cite{dakos2008slowing}, which involved linear interpolation to make the data equidistant and detrending with a Gaussian filter. Early warning signals were computed using a rolling window of 0.5.

\textbf{Construction activity in pre-Hispanic Pueblo societies:} This data is an annual-resolution time series of number of trees felled, a proxy for construction activity, in pre-Hispanic Pueblo societies~\cite{Tree_ring_Dates,scheffer2021loss,bocinsky2016exploration}. The data reveals five transitions in construction activity and subsequent distinct archaeological periods. We used the same data preprocessing and rolling window as in Scheffer et al. \cite{scheffer2021loss}. The time series were detrended using a Gaussian filter with a default bandwidth of 30 years, and early warning signals were computed with a default rolling window size of 60 years. The time series for the early and late PIII period were much shorter, therefore a smaller bandwidth (15 years) and rolling window (20 years) were used in these cases.

\bmhead{Data Availability}\label{sec_Data_Availability}

The chick heart data~\cite{Bury2023Predicting} are available at the GitHub repository \url{https://github.com/ThomasMBury/dl_discrete_bifurcation/tree/main/data/df_chick.csv}. The geochemical data \cite{sedimentary_archives} are available at the PANGAEA repository \url{https://doi.pangaea.de/10.1594/PANGAEA.923197}. The paleoclimate data \cite{Petit_Vostok} are available from the World Data Center for Paleoclimatology, National Geophysical Data Center, Boulder, Colorado (\url{http://www.ncdc.noaa.gov/paleo/data.html}). The tree-ring dates~\cite{Tree_ring_Dates} from the Southwestern United States can be accessed at the Digital Archaeological Record at \url{https://doi.org/10.6067/XCV82J6D7B}.

\bmhead{Code Availability}\label{sec_Code_Availability}

Code and instructions to reproduce the analysis are available at the GitHub repository \url{https://github.com/ZhiqinMa/surrogate_data_based_machine_learning}. The complete project data is archived on Zenodo at \url{https://zenodo.org/records/12562316}.

\bmhead{Acknowledgments}

This work was supported by the National Natural Science Foundation of China (Grant~Nos.~12265017 and 12247205), the Yunnan Fundamental Research Projects (Grant~Nos.~202201AV070003 and 202101AS070018), the Yunnan Ten Thousand Talents Plan Young \& Elite Talents Project, Yunnan Province Computational Physics, and Applied Science and Technology Innovation Team.

%\bmhead{Author contributions}
%Chunhua Zeng and Zhiqin Ma conceived the study. Zhiqin Ma developed the methodology. Zhiqin Ma performed the analysis. Chunhua Zeng provided resources. Chunhua Zeng provided project supervision. Zhiqin Ma wrote the first draft. All authors revised and commented on the manuscript.

\bmhead{Competing interests}
The authors declare no competing interests.

% \bmhead{Additional information}
\bmhead{Supplementary information} The online version contains supplementary material available at link (link building).


\bibliography{sn-bibliography}% common bib file

\end{document}